# Specific features of the kinetics of singlet fission in organic semiconductors. The effect of kinetic curves crossing.

A. I. Shushin

Institute of Chemical Physics, Russian Academy of Sciences,

119991, GSP-1, Kosygin Str. 4, Moscow, Russia

**Abstract**

Experimental investigations of magnetic field dependent kinetics of singlet fission (SF) processes in some organic semiconductors [i.e. splitting of excited singlet ($S_1$) state into a triplet exciton pair] have revealed the important specific feature of obtained kinetic curves, associated with decaying intensities $I(t)$ of fluorescence from $S_1$-state. Kinetic curves, measured in different magnetic fields, are found to cross each other. We show that this kinetic curves crossing (KCC) effect is a general feature of geminate condensed phase reactions, resulting from simple characteristic properties of kinetic schemes of processes. Specific features of the KCC-effect are analyzed in detail with some models of SF-processes.

## 1. Introduction

   Photochemical and photophysical condensed phase processes: radical pair (and polaron pair) recombination, singlet-fission (SF), triplet-triplet annihilation, etc., are actively investigated both experimentally and theoretically for very long time [1-15]. Intensive investigation of the kinetics of these processes is inspired by their great importance for applications [1-5].

   One of the most intensively studied processes is the SF in organic semiconductors, i.e. spontaneous splitting of the excited $S_1$-state into a pair of triplet (T) excitons, geminate annihilation of which essentially determines the short-time kinetics of decay of fluorescence $I(t)$ from the $S_1$-state in nanosecond time scales ($10^{-1}$ ns $< t < 10^2$ ns). The analysis of kinetic functions $I(t)$, called hereafter the kinetic curves, provides important information on characteristic features of mechanisms of SF-processes in various systems [2,5]. Note that the primary $S_1$-fission into TT-pair and subsequent geminate TT-annihilation are spin-selective processes and, therefore, are significantly affected by magnetic field [1-5]. The analysis of the magnetic field dependent SF-kinetics enable one to obtain additional information on SF-mechanisms [2,5].

   In this work we will discuss the interesting specific feature of SF-kinetics, in what follows denoted as the kinetic curves crossing (KCC), which is observed in a number of organic semiconductors [13-15] and is reproduced in some calculations [13,16,17]. The KCC-effect consists in crossing of kinetic curves $I(t)$, corresponding to different magnetic fields. This effect, however, is shown to be quite general property of the kinetics, characteristic not only of SF-processes, but also of a large number of other above-mentioned geminate reactions (and not only spin selective), which results from the special population conservation condition, satisfied in these reactions.

Specific properties of the KCC are analyzed as applied to SF-processes within two kinetic models: the simplified exponential model (of coupled first order processes) and the diffusion two-state model [18-20]. The evaluation of SF-kinetics in both models confirms general conclusions on the mechanism and conditions of the existence of the KCC-effect. Some other types of condensed-phase processes are proposed, in which the KCC-effect is expected to manifest itself as well.

## 2. General analysis.

In our work we will analyze some specific features of the kinetics of geminate condensed-phase reactive processes of general type. These features can easily and clearly be illustrated using (as an example) reactions, represented by the general chain kinetic scheme of $m$ kinetically coupled states

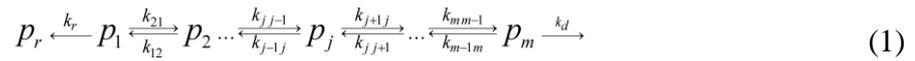
$$p_r \xleftarrow{k_r} p_1 \underset{k_{12}}{\overset{k_{21}}{\rightleftarrows}} p_2 \ldots \underset{k_{j-1\,j}}{\overset{k_{j\,j-1}}{\rightleftarrows}} p_j \underset{k_{j\,j+1}}{\overset{k_{j+1\,j}}{\rightleftarrows}} \ldots \underset{k_{m-1\,m}}{\overset{k_{m\,m-1}}{\rightleftarrows}} p_m \xrightarrow{k_d} \qquad (1)$$

in which $p_j(t)$ is the $j$-state population, ($j=1,\ldots,m$), $p_r(t)$ is the product state population, $k_{jj'}$ are rates of transitions $j \leftarrow j'$, $k_d$ is the rate of $m$-state decay, and $k_r$ is the rate of reaction in state 1.

In the analysis we will concentrate on the kinetics of geminate reactions (1), assuming the initial population of the reactive state 1: $p_1(t=0)=1$ and $p_{j\geq 2}(t=0)=0$. In general, the kinetics of reaction is described by the system of equations

$$\dot{p}_1 = k_{12}p_2 - (k_{21}+k_r)p_1; \quad \dot{p}_m = k_{m\,m-1}p_{m-1} - (k_{m-1\,m}+k_d)p_m;$$
$$\dot{p}_j = k_{j\,j+1}p_{j+1} + k_{j\,j-1}p_{j-1} - (k_{j+1\,j}+k_{j-1\,j})p_j, \quad (2 \leq j \leq m-1). \qquad (2)$$

These equations can be solved by the Laplace transformation $\tilde{p}_j(\varepsilon) = \int_0^\infty dt\, e^{-\varepsilon t} p_j(t)$. In our further consideration, however, we will not discuss the precise general solution, concentrating on the simple and evident relation $k_r \tilde{p}_1(0) + k_d \tilde{p}_m(0) = \int_0^\infty dt[k_r p_1(t) + k_d p_m(t)] = 1$, which expresses the population conservation in the process (1). Note, that in the important particular case of the absence of decay in the state $m$, i.e. for $k_d = 0$, this relation reduces to $\int_0^\infty dt\, p_1(t) = k_r^{-1}$.

The observable under study is the reaction kinetics, defined as the time dependent reaction flux

$$I(t) = k_r p_1(t) = \dot{p}_r(t). \qquad (3)$$

In the case $k_d = 0$ the population conservation results in obvious formula for the reaction yield $Y$:

$$Y = \int_0^\infty dt\, I(t) = 1, \qquad (4)$$

independent of values of rates $k_{jj'}$. The expression (4) allows one to draw some conclusions about the behavior of the kinetics $I(t)$ of the process.

The specific feature, which we will analyze, is the crossing of kinetic curves $I_i(t)$, corresponding to different sets of transition rates $\{k_{jj'}^{(i)}\}$ (but the same value of $k_r$) in kinetic equations (2). The existence of the KCC can easily be demonstrated using function

$$\phi_{ii'}(t) = I_i(t) - I_{i'}(t). \tag{5}$$

In the case $k_d = 0$, as follows from eq. (4), this function satisfies the relation

$$\tilde{\phi}_{ii'}(\varepsilon = 0) = \int_0^\infty dt\, \phi_{ii'}(t) = \int_0^\infty dt\, [I_i(t) - I_{i'}(t)] = 0. \tag{6}$$

Note that any smooth function $\phi_{ii'}(t) = I_i(t) - I_{i'}(t) \neq 0$, satisfying the relation (6) must, evidently, change its sign at least once and give positive and negative contributions to the integral (6) of the same absolute value. It is also worth noting that the change of $\phi_{ii'}(t)$-sign corresponds to the crossing of curves $I_i(t)$ and $I_{i'}(t)$ at times $t_k^\times$, satisfying equation $\phi_{ii'}(t_k^\times) = 0$.

**Thus, in the process (1), in the absence of non-reactive decay (in addition to the reaction in the state j = 1), the KCC-effect is observed, which shows itself in the crossing of kinetic curves $I_i(t)$, corresponding to different sets of rates $\{k_{jj'}^{(i)}\}$.**

To clarify this conclusion we will discuss its manifestation in some modified and generalized variants of the kinetic model (1), interesting for applications.

1) The model, often applied to describing SF-processes in organic semiconductors, is a particular case of the model (1), corresponding to $m = 3$. In the usually considered variant of the model the non-reactive decay is assumed to be absent and, therefore, in accordance with the above conclusion, SF kinetic curves, corresponding to different sets of rates $\{k_{jj'}^{(i)}\}$, must cross each other.

2) The conclusion, concerning KCC-effects, is formulated as applied to reactions of spinless particles. It is, nevertheless, valid in the case of spin-selective processes, since in this case the universal population conservation arguments used remain applicable as well.

3) The model (1) with infinitely large number of states *j* is often used to describe the effect of one-dimensional (1D) migration of particles on reaction kinetics (e.g. T-exciton migration effect on SF-kinetics). The above-formulated conclusion is also valid as applied to these reaction processes.

4) The proposed kinetic analysis with the model (1) can be essentially generalized by assuming diffusive motion of reacting particle at large interparticle distances (in the state *m*) [19]. The main conclusion on the existence of the KCC-effect turns out to be also valid in a number of these generalized variants of the model (1), based, for example, on approximations of free diffusion in spaces of low dimensionality $n \leq 2$. The fact is that for $n \leq 2$ diffusion is of recurrent type [21], implying that for diffusing particles the probability to return to the origin is unity, which ensures the validity of the population conservation condition and, thus, relations (4) and (6).

5) There are also some other diffusion models, in which relations (4) and (6) are fulfilled, resulting in validity of the conclusion on KCC-effects, in particular, the models with high reactivity in the reactive state (Sec. 4) and models of diffusion within potential wells and confined areas.

6) The above conclusion is, of course, still valid in the presence of fairly slow non-reactive decay of particles in states $j$. In general, the KCC-effect is expected to decrease with increasing the decay rate (Sec. 4) though this influence of the non-reactive decay can, however, be essentially reduced in the models, represented by the kinetic scheme (1) with a large number of intermediate states (large $m$) and slow transitions between these states (small rates $k_{jj'}$), which can efficiently "screen" the influence of fast non-reactive decay in remote states (with relatively large $j \sim m$) on the KCC-effect.

Below we illustrate the above-discussed KCC-effect, by considering SF-kinetics, as an example.

## 3. Kinetics of singlet fission processes.

Singlet fission is the important example of processes (1) (corresponding to $m=3$). Analysis of SF-kinetics will enable us to demonstrate specific features of the KCC-effect in SF-processes. The widely accepted formulation of SF-processes is represented by the scheme [2,5]

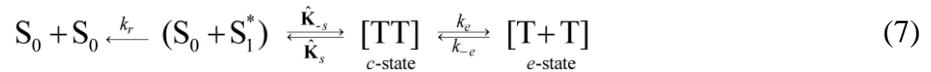

$$S_0 + S_0 \xleftarrow{k_r} (S_0 + S_1^*) \underset{\hat{K}_s}{\overset{\hat{K}_{-s}}{\rightleftarrows}} \underset{c\text{-state}}{[TT]} \underset{k_{-e}}{\overset{k_e}{\rightleftarrows}} \underset{e\text{-state}}{[T+T]} \qquad (7)$$

Here the primary SF-stage is the transition (with the rate $\hat{K}_{-s}$) from the initially excited state $(S_0 + S_1^*)$ into the intermediate [TT]-state ($c$-state) of a pair of coupled T-excitons (TT-pair). Evolution of [TT]-state is determined by TT-annihilation, escaping into [T+T]-state of separated excitons ($e$-state), and back capture into [TT]-state with rates $\hat{K}_s$, $k_e$ and $k_{-e}$, respectively. Note that the primary $S_1$-fission and TT-annihilation are spin-selective stages of SF-process with rates $\hat{K}_{-s}$ and $\hat{K}_s$, depending on the total TT-spin $\mathbf{S} = \mathbf{S}_a + \mathbf{S}_b$ of the pair of T-excitons, hereafter denoted as $a$ and $b$. Both processes are controlled by S-state of TT-pair (corresponding to $\mathbf{S}=0$): TT-pair is initially generated and then annihilate only in S-state.

SF-processes are accompanied by deactivation of $S_1^*$-state (with the rate $k_r$), resulting from radiation and radiationless transitions with rates $\kappa_r$ and $\kappa_r'$, respectively, (i.e. $k_r = \kappa_r + \kappa_r'$).

The observable under study is usually the normalized fluorescence: $\bar{I}(t) = I(t)/I(0)$, from $S_1^*$-state [13-16], determined by the $S_1^*$-state population $p_s(t)$, which in terms of the general model (1) corresponds to $p_1(t)$ ($p_s(t) \equiv p_1(t)$): $I(t) = \kappa_r p_s(t)$, (see Eq. (3)), so that $\bar{I}(t) = p_s(t)$.

In the presence of spin selectivity of primary $S_1$-fission and TT-annihilation stages SF-kinetics is essentially affected by the spin evolution of TT-pair, which is mainly controlled by the spin Hamiltonians of TT-pair in states [TT] and [T+T] (or $c$- and $e$-states): (below we assume that $\hbar = 1$)

$$\mathbf{H}_\nu = g\beta_B B(\mathbf{S}_a^z + \mathbf{S}_b^z) + \mathbf{H}_{T_\nu}^a + \mathbf{H}_{T_\nu}^b, \quad (\nu = c, e), \tag{8}$$

in which the first term is the Zeeman interaction of T-exciton spins with the magnetic field $\mathbf{B}$ and $\mathbf{H}_{T_\nu}^\mu$ are the ZFS-interactions in T-exciton $\mu$, $(\mu = a, b)$, (assumed to be the same in $c$ and $e$ states):

$$\mathbf{H}_{T_\nu}^\mu = D\left[(\mathbf{S}_{\nu_\mu}^{z_\mu})^2 - \tfrac{1}{3}\mathbf{S}_\mu^2\right] + E\left[(\mathbf{S}_{\nu_\mu}^{x_\mu})^2 - (\mathbf{S}_{\nu_\mu}^{y_\mu})^2\right] \; (\mu = a, b), \tag{9}$$

with $\mathbf{S}_{\nu_\mu}^{j_\mu}$ being the projection of the spin of the exciton $\mu$ (in the state $\nu = c, e$) along the eigenaxis $j_\mu$, $(j_\mu = x_\mu, y_\mu, z_\mu)$, of the ZFS-tensor [1,5].

TT-spin evolution, governed by spin-Hamiltonians $\mathbf{H}_\nu$, is described in the basis of $N = 9$ spin states of TT-pair, represented as products $|j_a j_b\rangle = |j_a\rangle |j_b\rangle$ of those $|j_\mu\rangle$, $(j = 1-3)$ for T-excitons. Hereafter it is convenient to use eigenstates of the Zeeman Hamiltonian $|j = 0, \pm\rangle$ (defined as $\mathbf{S}_\mu^{z_\mu}|j_\mu\rangle = j_\mu |j_\mu\rangle$) and eigenstates of the ZFS-interaction (7) $|j_\mu = x_\mu, y_\mu, z_\mu\rangle$ (defined by $\mathbf{S}_\mu^{j_\mu}|j_\mu = x_\mu, y_\mu, z_\mu\rangle = 0$) [4]. In particular, in these two bases $|S\rangle$-state of TT-pair is written as [5]

$$|S\rangle = \tfrac{1}{\sqrt{3}}(|0_a 0_b\rangle + |+_a -_b\rangle + |-_a +_b\rangle) = \tfrac{1}{\sqrt{3}}\sum_{j=x,y,z}|j_a j_b\rangle. \tag{10}$$

Noteworthy is that the rates $\hat{\mathbf{K}}_{-s}$ and $\hat{\mathbf{K}}_s$ are essentially determined by the operator $\hat{\mathbf{P}}_s = |S\rangle\langle S|$ of projection on $|S\rangle$-state ($\hat{\mathbf{K}}_{-s} \sim \hat{\mathbf{P}}_s$ and $\hat{\mathbf{K}}_s \sim \hat{\mathbf{P}}_s$ [1,5,16.]).

General precise evaluation of SF-kinetics (taking into account of TT-spin evolution and spin dependent rates $\hat{\mathbf{K}}_{-s}$ and $\hat{\mathbf{K}}_s$) reduces to solving the stochastic Liouville equation [1], which is a system of rather complicated equations for spin density matrices of TT-pairs in [TT] and [T+T] states [16], represented in the basis of vectors $|j_a j_b\rangle\langle j'_a j'_b|$ in the Liouville space [22].

In our work we are not going to discuss in detail this equation concentrating mainly on the analysis of the kinetic part of the problem. As to the spin evolution, it will be described with the approximate (but quite accurate [23]) Johnson-Merrifield approach (JMA),

3.1. *Johnson-Merrifield approach.*

Within the JMA the general problem reduces to solving simplified equations for vectors

$$\boldsymbol{\sigma} = \sum_{j_a j_b} \sigma_{j_a j_b} \| j_a j_b \rangle\rangle \text{ and } \boldsymbol{\rho} = \sum_{j_a j_b} \rho_{j_a j_b} \| j_a j_b \rangle\rangle \tag{11}$$

of spin-state populations in [TT] and [T+T] states in the basis $\| j_a j_b \rangle\rangle = |j_a j_b\rangle\langle j_a j_b|$ of diagonal elements of spin density matrices (populations),

$$\dot{p}_s = -(k_{-s} + k_r)p_s + (\mathbf{e}_s^+ \hat{K}_s \boldsymbol{\sigma})], \tag{12a}$$

$$\dot{\boldsymbol{\sigma}} = k_e \boldsymbol{\rho} - (\hat{W}_c + \hat{K}_s + k_e)\boldsymbol{\sigma} + N\hat{K}_{-s}\mathbf{e}_s p_s, \tag{12b}$$

$$\dot{\boldsymbol{\rho}} = -(\hat{W}_e + k_{-e})\boldsymbol{\rho} + k_e \boldsymbol{\sigma}, \tag{12c}$$

with the initial condition $p_s(t=0) = 1$, $\boldsymbol{\sigma}(t=0) = \boldsymbol{\rho}(t=0) = 0$. In these equations

$$\mathbf{e}_s = \tfrac{1}{N} \sum_{j_a j_b} \| j_a j_b \rangle\rangle \text{ and } \mathbf{e}_s^+ = \sum_{j_a j_b} \langle\langle j_a j_b \|, \text{ (with } \mathbf{e}_s^+ \mathbf{e}_s = 1), \tag{13}$$

are normalized equilibrium vector and adjoined one, in which $N = 9$ is the number of TT-spin states. Matrices $\hat{K}_{-s}$ and $\hat{K}_s$ are the JMA-representations of rate matrices $\hat{\mathbf{K}}_{-s}$ and $\hat{\mathbf{K}}_s$, written as

$$\hat{K}_\alpha = k_\alpha \hat{P}_s, (\alpha = -s, s), \text{ with } \hat{P}_s = \sum_{j_a j_b} C^S_{j_a j_b} \| j_a j_b \rangle\rangle\langle\langle j_a j_b \| \tag{14}$$

where $\hat{P}_s$ is the matrix of weights $C^S_{j_a j_b} = |\langle S | j_a j_b \rangle|^2$ of S-state in states $|j_a j_b\rangle$ (with $\sum_{j_a j_b} C^S_{j_a j_b} = 1$). The rate matrices $\hat{W}_c$ and $\hat{W}_e$ describe spin-lattice relaxation in states $c$ and $e$.

In general, one can solve eqs (12a)-(12c) by Laplace transformation in time [defined as $\tilde{p}(\varepsilon) = \int_0^\infty dt\, e^{-\varepsilon t} p(t)$ for any $p(t)$] and, thus, obtain general expression for $\tilde{p}_s(\varepsilon)$ [16].

In our work, however, we will restrict ourselves to the illustrative analysis of SF-kinetics, previously studied in amorphous rubrene films[13] in limiting cases of weak and strong magnetic fields ($B \ll B_T = \|\mathbf{H}^\mu_{T_\nu}\|/(g\beta_B) \sim D/(g\beta_B)$ and $B \gg B_T$, respectively), considering quite realistic systems, in which spin-lattice relaxation T-excitons is fairly fast [16].

In these systems:

(1) *In the case of weak* $B \ll B_T = \|\mathbf{H}^\mu_{T_\nu}\|/(g\beta_B)$ the fast spin-lattice relaxation, for which $\|\hat{W}_{c,e}(B \ll B_T)\| = w^<_{c,e} > (k_s/N_<)$ (where $N_< = N = 9$ is the total number of TT-spin states), averages the spin dependence of primary $S_1$-fission and TT-annihilation rates, resulting in equal reactivity of all $N_< = N = 9$ spin states $\|j_a j_b\rangle\rangle$ of TT-pairs and, thus, in expressions [16]

$$\hat{K}_\alpha = \kappa^<_\alpha \hat{E}_<, (\alpha = -s, s) \text{ with } \kappa^<_\alpha = \tfrac{1}{N_<} k_\alpha = \tfrac{1}{9} k_\alpha \tag{15}$$

and $\hat{E}_< = \sum_{j_a j_b} \| j_a j_b \rangle\rangle\langle\langle j_a j_b \|$ in the unity matrix in the space of 9 reactive spin states.

(2) *In the case of strong* $B \gg B_T$ the magnetic field can significantly affect the spin-lattice relaxation resulting in strong decrease of relaxation rates [16]. The fact is that for $B \gg B_T$ there are only $N_> = 3$ nearly degenerate states (of TT-pair), $|1\rangle \approx |0_a 0_b\rangle$, $|2\rangle \approx |+_a -_b\rangle$, and $|3\rangle \approx |-_a +_b\rangle$ with equal singlet character (and splitting $\omega \sim D/[D/(g\beta B)] \ll D$ [16] for non-equivalent T-excitons). In strong B, for which $\Omega_B = g\beta_B B \gg \tau_c^{-1} > g\beta_B B_T$ (where $\tau_c$ is the correlation time of $\mathbf{H}^\mu_{T_\nu}$-fluctuations resulting from jumps of T-excitons over non-equivalent spatial states [16]), the rates $w_{c,e}^{>}$ of relaxation transitions between these 3 spin states and other 6 spin states are reduced because of large Zeeman-interaction-induced splitting between these two groups of states: $w_{c,e}^{>} = w_{c,e}^{<}/(\Omega_B \tau_c)^2 \ll w_{c,e}^{<}$ [5,16]. In this limit primary $S_1$-fission and TT-annihilation rate matrices $\hat{K}_{-s}$ and $\hat{K}_s$ describe SF-processes only in $N_{>} = 3$ ($|1\rangle - |3\rangle$) equally reactive states:

$$\hat{K}_\alpha = \kappa_\alpha^{>} \hat{E}_{>}, \; (\alpha = -s, s) \text{ with } \kappa_\alpha^{>} = \tfrac{1}{N_{>}} k_\alpha = \tfrac{1}{3} k_\alpha, \qquad (16)$$

where $\hat{E}_{>} = \sum_{j=1-3} \| j \rangle\!\rangle\langle\!\langle j \|$ is the unity matrix in the space of 3 reactive spin states $\| j \rangle\!\rangle = | j \rangle\langle j |$.

**In conclusion, presented formulas show that in the considered example of SF-processes the magnetic field effect reduces to the change of the number $N_{SF}$ and reactivity $\kappa_\alpha$ ($\alpha = -s, s$) of TT-spin states, equally involved in the process, from $N_{SF} = N_{<} = 9$ and $\kappa_\alpha = \kappa_\alpha^{<} = k_\alpha/9$ (for weak fields $B \ll B_T$) to $N_{SF} = N_{>} = 3$ and $\kappa_\alpha = \kappa_\alpha^{>} = k_\alpha/3$ (for strong fields $B \gg B_T$)** [16]**.**

Solution of equations (12a)–(12c) in these two limits yields universal formula for $\tilde{p}_s(\varepsilon)$:

$$\tilde{p}_s(\varepsilon) = \{\varepsilon + k_{rs} - (k_{-s}\kappa_s)/[\varepsilon + \kappa_s + K_e(\varepsilon)]\}^{-1}, \qquad (17)$$

with $k_{rs} = k_r + k_{-s}$ and the effective rate of escape from the [TT]-state

$$K_e(\varepsilon) = k_e - k_e k_{-e}(\varepsilon + k_{-e})^{-1} = \varepsilon k_e (\varepsilon + k_{-e})^{-1}. \qquad (18)$$

In general, the expression (17) is still fairly cumbersome, containing a large number of parameters, and is not quite convenient for the analysis of the KCC-effect, i.e. crossing of curves $p_s(t)$, corresponding to different values of the magnetic field. To simplify and somewhat generalize the analysis we will consider two particular variants of the process (7).

3.2. *Particular examples of the kinetic model of SF-processes.*

3.2.1. *The simplified exponential model.*

In the simple variant of the model (7), which we will call the simplified exponential model, we assume that the decay of [TT]-state is irreversible, i.e. $k_{-e} = 0$, so that $K_e(\varepsilon) = k_e$ and

$$\tilde{p}_s(\varepsilon) = [\varepsilon + k_{rs} - (k_{-s}\kappa_s)/(\varepsilon + k_{es})]^{-1}, \qquad (19)$$

with $k_{rs} = k_r + k_{-s}$ and $k_{es} = k_e + \kappa_s$. Therefore

$$p_s(t) = Z_{+} \exp(-\varepsilon_{+} t) + Z_{-} \exp(-\varepsilon_{-} t), \qquad (20)$$

where $\varepsilon_{\pm}$ are the roots of equation $(k_{rs} - \varepsilon)(k_{es} - \varepsilon) = k_{-s}\kappa_s$:

$$\varepsilon_{\pm} = k_{+} \pm \Delta \text{ and } Z_{\pm} = (\Delta \mp k_{-})/(2\Delta) \qquad (21)$$

with

$$k_{\pm} = \tfrac{1}{2}(k_{rs} \pm k_{es}) \quad \text{and} \quad \Delta = \sqrt{k_{-}^2 + \kappa_s k_{-s}} . \tag{22}$$

*3.2.2. The diffusion two-state model.*

The diffusion two-state model of SF-processes [16] can be treated as a generalization of the exponential model (12a)-(12c). In the two-state model the first stage $(S_0 + S_1^*) \leftrightarrow [TT]$ is considered as a first order reversible process, while the second stage $[TT] \leftrightarrow [T+T]$ is described in the two-state approach, developed earlier for analyzing the diffusive escape of a particle from a potential well [18-20]. Within this approach the evolution of *geminate* TT-pairs is represented as transitions between two states: [TT]-state of interacting T-excitons and [T+T]-state of separated T-excitons, undergoing diffusive migration.

The diffusion two-state model predicts simple universal analytical formula for the Laplace transform of SF-kinetics $\tilde{p}_s(\varepsilon)$ valid in the case of 1D and 3D diffusion in [T+T]-state [16]

$$\tilde{p}_s(\varepsilon) = \{\varepsilon + k_{rs} - (k_{-s}\kappa_s)/[\varepsilon + \kappa_s + \overline{K}_e(\varepsilon)]\}^{-1} , \tag{23}$$

similar to eq. (17), but with effective rate of diffusion escape from [TT]-state [16]

$$\overline{K}_e(\varepsilon) = k_e + \kappa_e (\varepsilon/\kappa_e)^{1/2} . \tag{24}$$

In formulas (24) $k_e$ is the steady state rate of diffusion controlled [TT]-state decay: $[TT] \to [T+T]$, and $\kappa_e$ is the rate parameter, characterizing transient (reencounter) part of the decay rate [16,19], determined by specific features of TT-interaction in [TT]-state (*c*-state) and diffusive migration in [T+T]-state. Noteworthy is that values of the rates depend on the space dimensionality *n* [19,20]. In particular, *for isotropic* 3D *diffusion* (*n* = 3) $k_e \neq 0$, *while for* 1D *diffusion* (*n* = 1) $k_e = 0$ [19,20].

Note that just the relation $k_e = 0$ for *n* = 1 ensures the fulfillment of the population conservation relation $\tilde{p}_s(0) = \int_0^\infty dt\, p_s(t) = k_r^{-1}$, which is the analog of eq. (4) in terms of function $p_s(t)$.

## 4. Results and discussion.

Above-discussed models of SF-processes enable one to demonstrate and analyze the KCC-effect, i.e. crossing of kinetic dependences $I_i(t)$, corresponding to different sets of transition rates $\{k_{jj'}^{(i)}\}$ (except for $k_r$) in kinetic equations (1). Here we will illustrate the KCC-effect within two models (simplified exponential and diffusion two-state ones), considered in Sec. 3, in limits of weak and strong magnetic field.

It is worth recalling that, in accordance with the general analysis in Sec. 3, for the particular variant of SF-process under study these magnetic field limits manifest themselves only in the values of characteristic rates $\kappa_\alpha$, $(\alpha = -s, s)$, and the number of spin states, involved in the process, as it follows from eqs. (15) and (16).

Note also that in results of calculations, presented below, the set of rates of considered models are represented in the form of dimensionless parameters

$$z_r = k_r / k_{rs}, \quad z_e = k_e / k_{rs}, \quad \text{and} \quad z_s = \kappa_s / k_{rs}. \tag{25}$$

For the diffusion two-state model the parameter $\bar{z}_e = \kappa_e / k_{rs}$ should also be added (see below).

In our further analysis we will take $z_r, z_s$, and $\bar{z}_e$ of values close to those used for treating SF-kinetics in realistic systems [16], and three different $z_e$–values to illustrate the origin of the KCC.

### 4.1. *The simplified exponential model of SF-processes.*

The simplified exponential model (19) is very suitable for the analysis of the KCC-effect in SF-processes. The behavior of SF-kinetics, evaluated for three sets of rate parameters (25): one set with $z_e = 0$, and two sets with $z_e \neq 0$ and values of $z_e$ and $z_s$, corresponding to two different relations between $k_e$ and $\kappa_s^\gamma$ [where $\gamma = >, <$ ; (see eqs (15) and (16))]: $k_e < \kappa_s^\gamma, k_r$ and $k_e > \kappa_s^\gamma, k_r$, are shown in Fig. 1. In agreement with results of Sec. 3 the KCC-effect appears only for small rates $k_e$, ensuring the weakness of violation of the population conservation relation (4). Moreover, at $k_e = 0$ the KCC-effect shows itself most distinctively, as expected. It is also seen from Fig.1 that the KCC time $t_c^\times$ increases with increasing the rate $k_e$ resulting in disappearance of the KCC at large $k_e$.

Some additional arguments in favor this qualitative conclusion can be obtained by more detailed analysis of asymptotic short and long time behavior of SF-kinetics $p_s(t)$, predicted by the model.

(1) *Short time dependence of $p_s(t)$.* The dependence of $p_s(t)$ at short times is determined by $\tilde{p}_s(\varepsilon)$ (19) at large $\varepsilon$: $\tilde{p}_s(\varepsilon) \approx [\varepsilon + k_r + (k_{-s} k_e)/(\kappa_s + k_e)]^{-1}$, which predicts the dependence

$$p_s(t) \approx e^{-w_- t} \quad \text{with} \quad w_- \approx k_r + (k_{-s} k_e)/(\kappa_s + k_e). \tag{26}$$

It is seen that the rate $w_-$ of exponential short time decrease of $p_s(t)$ decreases with increasing $\kappa_s$.

(2) *Long time dependence of $p_s(t)$.* At long times SF-kinetics $p_s(t)$ is determined by the most slowly decreasing second term in eq. (20): $p_s(t) \sim \exp(-\varepsilon_- t)$. Moreover, the effect of crossing of $p_s(t)$-curves for different $\kappa_s$ is mainly controlled by specific features of $\varepsilon_-(\kappa_s)$-dependence. For example, in the limit $\kappa_s, k_{-s} \gg k_e, k_r$ $\varepsilon_-(\kappa_s)$ can approximately be estimated by formula

$$\varepsilon_-(\kappa_s) \approx [k_e + k_r (\kappa_s / k_{-s})]/[1 + (\kappa_s / k_{-s})], \tag{27}$$

which shows, in particular, that for small $k_e \ll k_r(\kappa_s/k_{-s})$ the function $\varepsilon_-(\kappa_s)$ increases with the increase of $\kappa_s$: $\varepsilon_-(\kappa_s) \sim \kappa_s$, thus resulting in faster long time decrease of $p_s(t)$ for larger $\kappa_s$, i.e. in crossing of $p_s(t)$ for different $\kappa_s$.

4.2. *The diffusion two-state model of SF-processes.*

The above-described more realistic diffusion two-state model (Sec. 3.2) allows us to provide some additional examples of manifestations of the KCC. Figure 2 displays SF-kinetics for three sets of rate parameters (25) as well: one set with $z_e = 0$, and two sets with $z_e \neq 0$ and values of $z_e$ and $z_s$, corresponding to two different relations between $k_e$ and $\kappa_s^\gamma$ [where $\gamma =>,<$ ; (see eqs (15) and (16))]: $k_e < \kappa_s^\gamma, k_r$ and $k_e > \kappa_s^\gamma, k_r$, similar to those applied above in the analysis of the simplified exponential model. In agreements with predictions of the simplified exponential model, the two-state model indicates the appearance of the KCC-effect only for small escape rates $k_e$ [which ensure the weakness of the violation of the population conservation relation (5)], with most pronounced manifestation of the KCC-effect observed at $k_e = 0$. Note also that, as in the simplified exponential model, the KCC time $t_c^\times$ increases as the value of the rate $k_e$ is increased finally resulting in the disappearance of the KCC.

(1) *Short time dependence of $p_s(t)$*. The short time dependence of $p_s(t)$, predicted by the diffusion two-state model, is similar to that of the simplified exponential model (20) because of the similarity of description of early (conventional first order) SF-stages. This means that the rate $w_-(\kappa_s)$ of exponential SF-decay, $p_s(t) \approx e^{-w_- t}$, decreases with the increase of $\kappa_s$ [see eq. (26)].

(2) *Long time dependence of $p_s(t)$*. As to the two-state-model predicted long-time part of kinetics $p_s(t)$, it is of inverse power type one [19,20], i.e. essentially different from that obtained in the exponential model (20). The $\kappa_s$ dependence of this long time dependence can be obtained by expansion of $\tilde{p}_s(\varepsilon)$ [eqs (23), (24)] in small $\varepsilon$ and subsequent inverse Laplace transformation:

$$p_s(t) \sim \frac{p_+(\kappa_s)}{(\kappa_e t)^{3/2}} \quad \text{with} \quad p_+(\kappa_s) = \frac{\kappa_s \kappa_e}{(\alpha_s k_e + \beta_s \kappa_s)^2}, \tag{28}$$

$\alpha_s = 1 + k_r/k_{-s}$ and $\beta_s = k_r/k_{-s}$. It is easily seen from eq. (28) for small $k_e$ the function $p_+(\kappa_s) \sim 1/\kappa_s$ decreases with the increase of $\kappa_s$, thus resulting in the KFC in SF-processes in agreement with above-presented general analysis (Sec. 2). It is important to note that in the case of 1D diffusion $k_e = 0$ and, therefore, $p_+(\kappa_s) \sim \kappa_s^{-1}$ for any value of $\kappa_s$, which means that the KCF-effect is always observed in SF-kinetics, as it follows from results of the analysis in Sec. 2.

## 5. Conclusions.

Recent experimental investigations of magnetic field dependent SF-processes in molecular organic semiconductors have revealed interesting characteristic property of the kinetics of these processes at short times ($<10^{-8}$ s ): the kinetics curves, i.e. time dependent intensities $I_i(t)$ of fluorescence (from the excited S$_1$-state), at different magnetic fields [13-15] cross each other.

In this work the effect of kinetic curves crossing (KCC) is shown to be a general phenomenon, which can be observed in not only in SF-processes, but also in some other reactions. It results from the conservation of population of excited states, involved in processes under study (Sec. 2). The origin of the KCC-effect is demonstrated within generalized kinetic scheme (1), though the effect can be realized under even more general assumptions, as mentioned in Sec. 2.

The validity of predictions as well as the characteristic behavior of the KCC-effect is illustrated within two models of SF-kinetics: the simplified exponential model (7) and recently developed diffusion two-state model (23), which allows for the accurate treatment of the effect of T-exciton diffusion on SF-kinetics [16]. The results of calculations in both models confirm the existence of the KCC-effect for parameters of models, predicted by the general analysis of Sec. 2.

In conclusion, it is worth adding some comments on processes, in which this effect can be observed.

(1) Our analysis concerned reactions whose kinetic scheme (1) is of "1D" geometry. The KCC-effect, however, can also be observed in processes of more complicated "2D" and "3D" geometries described by kinetic schemes of branched structure, in which the occurrence of the KCC-effect is ensured, due to the finite "size" of kinetic schemes, representing the processes and by the absence of any population decay channels apart from the reaction channel.

(2) In the proposed analysis the rates { $k_{jj'}$ } of transitions between kinetic states are assumed to be independent of time. In reality, however, the KCC-effect is also expected to be observed in the reactions of type of (1) with time dependent transition rates { $k_{jj'}(t)$ }, as it follows from general analysis of Sec. 2. Moreover, the rates $k_{jj'}$ can depend on populations $\mathbf{p} = \{p_j\}, (2 \leq j \leq m-1)$: $[k_{jj'} \equiv k_{jj'}(\mathbf{p})]$, resulting in nonlinear kinetic equations (2). These equations, nevertheless, predict the validity of population conservation relations (4) and (6) and thus existence of the KCC-effect.

(3) The analysis, presented in Sec. 2, does not allow to obtain the number of KCC times $t_k^\times$. For example, in above-considered simple models there is only one KCC time. However, in more complex models, describing multistage processes with significantly different timescales, the number of $t_k^\times$ can be fairly large (corresponding to the number of characteristic times of process stages).

(4) In our analysis we have assumed the presence of non-reactive decay only in the state $j = m$. In general, however, the decay in other states ($j = 2,\ldots,m\text{-}1$) is also possible (see Sec. 2). The additional decay in these states is expected to lead to extra reduction of the KCC-effect, making it less pronounced and may be resulting in subsequent disappearance of the KCC at very large decay rates. The form of the reduction, however, depends on the relation between rates $k_{jj'}$ as well as the relation between rates $w_j$ of decay of populations $p_j$ in states $j$ (as it is mentioned in Sec. 2).

**Acknowledgements**

This work was supported by the Russian state grant AAAA-A17-117032750003-9 # 0082-2014-0019 and the Russian Foundation for Basic Research [No. 16-03-00052].

# Figures

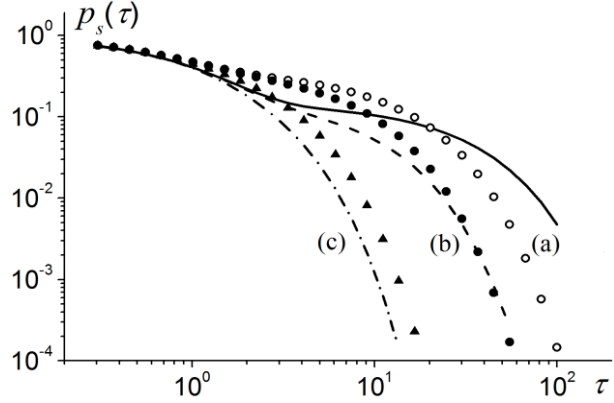

**Figure 1**. The dependence of the normalized SF-kinetic function $p_s(\tau) = \bar{I}(\tau) \equiv I(\tau)/I(0)$ (Sec. 3) on dimensionless time $\tau = k_{rs}t$, calculated within the simplified exponential model (19) for $z_r = 0.2$; two values $z_s = z_s^< = 0.2$ (lines) and $z_s = z_s^> = 0.6$ (points) [eqs (15), (16) and (25)], and three values of $z_e = k_e/k_{rs}$: (a) $z_e = 0$, and $z_s = z_s^<$ (full line), $z_s = z_s^>$ (open circles); (b) $z_e = 0.1$, and $z_s = z_s^<$ (dashed line), $z_s = z_s^>$ (black circles); (c) $z_e = 0.8$, $z_s = z_s^<$ (dash-dot line), $z_s = z_s^>$ (triangles).

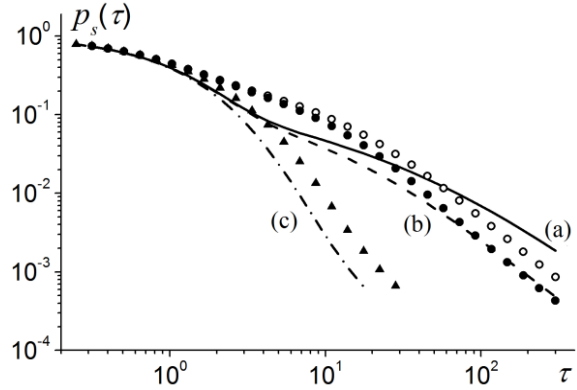

**Figure 2**. The dependence of the normalized SF-kinetic function $p_s(\tau) = \bar{I}(\tau) \equiv I(\tau)/I(0)$ (Sec. 3) on dimensionless time $\tau = k_{rs}t$, calculated within the diffusion two-state model (23) for $z_r = 0.2$; $\bar{z}_e = \kappa_e/k_{rs} = 0.2$; two values $z_s = z_s^< = 0.2$ and $z_s = z_s^> = 0.6$ [eqs (15), (16) and (25)], and three values of $z_e = k_e/k_{rs}$: (a) $z_e = 0$, and $z_s = z_s^<$ (full line), $z_s = z_s^>$ (open circles); (b) $z_e = 0.05$, and $z_s = z_s^<$ (dashed line), $z_s = z_s^>$ (black circles); (c) $z_e = 0.8$, and $z_s = z_s^<$ (dash-dot line), $z_s = z_s^>$ (triangles).

14